\documentclass[aip]{revtex4-1}
\usepackage{graphicx}
\usepackage[version=4,arrows=pgf-filled,
textfontname=sffamily,
mathfontname=mathsf]{mhchem}
\usepackage{bm}
\usepackage{fancyhdr}
\usepackage{amsmath}
\usepackage{amsfonts}
\usepackage{amsbsy}
\usepackage{amssymb}
\usepackage[mathscr]{eucal}
\usepackage{color}
\usepackage{soul}
\usepackage{mathtools}
\usepackage[english]{babel}
\usepackage{verbatim}
\let\originalselectlanguage\selectlanguage
\renewcommand{\selectlanguage}[1]{%
  \def\langinput{#1}%
  \ifnum\pdfstrcmp{\langinput}{en}=0
    \originalselectlanguage{english}%
  \else\ifnum\pdfstrcmp{\langinput}{eng}=0
    \originalselectlanguage{english}%
  \else
    \originalselectlanguage{#1}%
  \fi\fi
}

\begin{document}
\title{Speed limits on biomolecular processes from fundamental physical constants}
\author{Dmitrii E Makarov }
\email{makarov@cm.utexas.edu}
\affiliation{Department of Chemistry, University of Texas at Austin, Austin, TX, 78712 }
\affiliation{Oden Institute for Computational Engineering and Sciences, University of Texas at Austin, Austin, TX, 78712}

\begin{abstract}
    Many of the timescales of life have speed limits set  by quantum-mechanical constraints along with non-fundamental quantities, such as the temperature of the environment, which are however bounded by anthropic considerations.  Here, some  of such speed limits are examined, including those for the rates of elementary chemical reactions and biomolecular folding. Limitations of simple back-of-envelope estimates are also discussed.  
\end{abstract}
\maketitle

Why is the shortest lifespan for a living organism a few minutes and not a microsecond? Arguably, this is (at least partly) because essential biochemical transformations, which are elementary steps constituting more complex phenomena such as replication, take time. But what sets the limit on how fast elementary chemical steps can take place?         

In a beautiful discussion\cite{weisskopf_atoms_1975}, Victor Weisskopf showed how to estimate various microscopic and macroscopic scales starting from fundamental physical constants. He starts with estimating the size of the hydrogen atom,  moves on to estimating the strength of cohesive interactions within a material and proceeds to predict the height of the tallest possible mountain on Earth entirely on the basis of  fundamental physical constants. Several recent works\cite{press_mans_1980,trachenko_minimal_2020,mehta_what_2025} applied Weisskopf-style analysis to estimate the dynamical timescales in physics and biophysics. The purpose of this note is to extend some of those ideas to understand the timescales of biomolecular phenomena.     

The rate of an elementary chemical transformation from one chemical species to another is set by the Arrhenius law,
\begin{equation} \label{Arrhenius}
    k=\nu e^{-\frac{E_a}{k_B T}}
\end{equation}
When the trasformation involves breaking or formation of a covalent bond, the activation energy $E_a$ is comparable in scale to the strength of a chemical bond -- let us call that scale $E_b$. For gas-phase reactions, the prefactor $\nu$ is often  of the order of  $\sim 10^{13} s^{-1}$, while it could be significantly lower for reactions in solution.  The prefactor sets the speed limit of a chemical reaction\footnote{The term ``speed limit'' is understood here in a somewhat loose sense. In the quantum information field, speed limits of quantum processes are usually defined more precisely in terms of rigorous inequalities\cite{deffner_quantum_2017}. Interestingly, the problem considered here is closely connected to and can be unified with that of quantum speed limits\cite{del_campo_quantum_2026}}: with the energy barrier extrapolated to zero, the reaction could not go faster than $\nu$. The Arrhenius law can also be used to describe the rate of protein folding or unfolding, but its speed limit $\nu$ is about 5 orders of magnitude lower: does this difference have something to do with the chemistry of proteins (i.e., polypeptide molecules), or could we anticipate this result  on more general grounds?

\subsection*{How fast can a covalent bond break?} This timescale is set by the prefactor $\nu$. Following Eyring, many chemistry textbooks state that its value is given by
\begin{equation}
    \nu \equiv \nu_{Ey} = \frac{k_B T}{2\pi \hbar }\approx 6\times 10^{12} s^{-1}\sim 10^{13} s^{-1}, 
\end{equation}
where $T\approx 300K$ is the typical temperature of our environment. We will call $\nu_{Ey}$ the Eyring frequency and its inverse $\tau_{Ey}=\nu_{Ey}^{-1}$ the Eyring time, which is between femtoseconds and picoseconds. But modern transition state theory\cite{pollak_reaction_2005,peters_reaction_2017,makarov_single_2015} gives a different estimate for this prefactor: it is the oscillation frequency along the bond that is broken\footnote{Or, more precisely, along a properly chosen ``reaction coordinate''; for the purpose of this note, which only deals with orders of magnitude, a ``molecular vibration frequency'' is close enough.}, 
\begin{equation} \label{prefactor}
    \nu \approx \frac{\omega}{2\pi}
\end{equation}
Interestingly, molecular vibration frequencies are usually comparable to the Eyring frequency, making Eyring's estimate reasonable despite the fact that the standard transition state theory is a {\it classical} theory which, which should have no place for Planck's constant\footnote{We note that even quantum transition theory does not have $\nu_{Ey}$ as a prefactor\cite{benderskii_chemical_1994}}! 

This remarkable coincidence deserves further investigation. We proceed in Weisskopf's style: A typical vibration frequency can be estimated as 
$$
\omega = \sqrt{\kappa/m},
$$
where $m$ is an (effective) molecular mass that vibrates and 
\begin{equation} \label{stiffness}
    \kappa \sim \frac{E_b}{a^2}.
\end{equation}
is a bond's spring constant. Here $E_b$ describes the energy scale of molecular bonds and
$$
a\approx \frac{\hbar^2}{m_e e^2} 
$$
is the Bohr radius (with $m_e$ denoting the electron mass), which sets a length scale for molecular interactions. Moreover, the scale of the bond energy is roughly set by the dissociation energy of the hydrogen atom,   
\begin{equation} \label{Rydberg}
    1 \text{Ry} = \frac{\hbar^2}{2 m_e a^2}.
\end{equation}
Typically, $E_b$ is lower than (but within an order of magnitude of) $1 \text{Ry}$; if we wish to refine our estimate we write 
$$
E_b = \xi \,\text{Ry}
$$
with $\xi<1$ being typically between 0.1 and 1. 
The approximate equality
\begin{equation} \label{Eyring equals frequency}
\nu_{Ey}\approx\omega/{2\pi}
\end{equation}
then implies that
$$
\frac{k_B T}{E_b} \approx \sqrt{\frac{m_e}{m}} \sqrt{\frac{2}{\xi}} = \sqrt{\frac{m_e}{m_p}} \left\{\sqrt{\frac{2}{\xi}} \sqrt{\frac{m_p}{m}} \right\},
$$
where $m_p$ is the proton mass. The quantity in the curly brackets here is a product of two factors, the first being larger than 1 and the second smaller then 1 (but both $O(1)$). If we assume that this quantity is on the order of $1$ then we write, approximately,
\begin{equation}  \label{thermal energy}
\mathcal{A}  \equiv \frac{E_b}{k_B T} \sim  \sqrt{\frac{m_p}{m_e}}, 
\end{equation}
where we have defined the ``Arrhenius ratio'' 
$$\mathcal{A}=\frac{E_b}{k_B T}\sim \frac{\hbar^2}{m_e a^2 k_B T}$$
The order-of-magnitude similarity stated by Eq.~\ref{thermal energy} is indeed true; for example, if $E_b \sim 1 \text{eV}$ then the 
left-hand side of Equation \ref{thermal energy} is $\approx 40$ and the right-hand side is $\approx 43$! Therefore, the order-of-magnitude agreement between the Eyring speed limit $\nu_{Ey}$ and typical molecular vibrational frequencies appears to be a numerological accident, i.e., the similarity of the Arrhenius ratio (i.e., the ratio of bond energy and thermal energy) and the square root of the ratio of the proton and electron masses. 

The thermal energy $k_B T$ is not a fundamental constant, and we could imagine living on a planet where this ratio would be quite different, even if the order-of -magnitude estimate for $\mathcal{A}$ does not significantly change even on Mars or Venus. We may nevertheless ask whether there is any reason why this ratio has the value it does. First we notice that this is roughly the ratio entering the exponent of Eq.~\ref{Arrhenius}. If it were much larger then the rate $k$ would be to small, and no chemical reactions could happen at all. Thus we could estimate an upper limit on the Arrhenius ratio by demanding that 
$$ 
k\sim \nu e^{-\mathcal{A}} > \frac{1}{\text{age of the Universe}}.
$$
Using $\sim 10^{17} s$ for the age of the Universe and assuming $\nu \sim 10^{13} s^{-1}$, we find $\mathcal{A}<70$. On the other hand, it would be disruptive to our existence if the covalent bonds linking the chain molecules of life were to frequently dissociate spontaneously. If we demand that those bonds remain stable, say, over an hour we get a lower bound on the Arrhenius ratio
$$ 
k\sim \nu e^{-\mathcal{A}} < \frac{1}{\text{hour}},
$$
which gives $\mathcal{A}>38$. Of course this anthropic argument\cite{barrow_constants_2004} is cheating from this paper's point of view, as it relies on our knowledge about a typical lifespan. Nevertheless, the existence of life as we know it necessitates 
$$
38<\mathcal{A}<70,
$$
and so the approximate equality of the Arrhenius ratio and the square root of the ratio of the proton and electron masses, Eq.\ref{thermal energy}, no longer appears like a numerological coincidence!

\subsection*{Randomness, diffusion, and dissipation in molecular dynamics} Chemical processes of life take place in solution. Consider the motion of a small molecule across its aqueous environment.  It is highly erratic, as the molecule incessantly bumps into the surrounding molecules. A useful mathematical model of this motion is borrowed from the theory of Brownian motion: 
%Eight decades ago, Kramers proposed\cite{kramers_brownian_1940} to treat vibrational degrees of freedom of molecules in the same manner, by assuming that the motion along a molecular degree of freedom, such as a vibrational coordinate, 
According to the Langevin equation,
\begin{equation}\label{Langevin}
    m \ddot{x} = -U'(x)-\gamma \dot{x} + f(t),
\end{equation}
the molecule of mass $m$, whose position is characterized by a coordinate $x$, is subjected to a random  force $f(t)$  that describes random kicks received from the surroundings (solvent) and to a friction force $\gamma \dot{x}$  that is proportional to the molecule's velocity. In addition, we have introduced a potential (of mean force)  $U(x)$ describing other external forces on the molecule. 

A ``spherical cow'' model of a molecule in solution describes it as a spherical Brownian particle of radius $R$ in a liquid of bulk viscosity $\eta$. If so, the friction coefficient can be calculated using the Stokes formula, 
\begin{equation} \label{Stokes}
    \gamma = 6\pi \eta R. 
\end{equation}
Although one could object to applying the Stokes formula to microscopic objects such as molecules, this approximation is not unreasonable and is employed routinely in coarse-grained simulations of biomolecules and in theories of polymer dynamics. 

Let us first focus on the motion of a free molecule in zero potential, $U(x)=0$. At short timescales the Langevin equation (as well as physical intuition) predicts that the molecule will move with a constant velocity, while at longer timescales its motion will look like a random walk or diffusion. The crossover between these two regimes occurs at a ``velocity'' timescale 
$$
\tau_{vel} \approx \frac{m}{\gamma}=\frac{m}{6 \pi \eta R}, 
$$
which is the time it takes the molecule to forget its own velocity\cite{robert_zwanzig_nonequilibrium_2001}. At $t\gg \tau_{vel}$, the motion of the molecule is diffusion, with the mean square displacement growing linearly in time, 
$$
\langle x^2(t) \rangle = 2D t,
$$
where the diffusivity $D$ is given by the Stokes-Einstein formula
$$
D=\frac{k_B T}{\gamma}=\frac{k_B T}{6 \pi \eta R}.
$$
For a small molecule we expect that the radius  $R$ should be on the order of a molecular size, which we will estimate as the Bohr radius $a$.   Trachenko and Brazhkin (TB)  proposed a quantum limit on the minimum possible value of the viscosity of a liquid given by\cite{trachenko_minimal_2020}
\begin{equation} \label{Trachenko_Brazhkin}
    \eta \equiv \eta_{TB}\sim \frac{\hbar}{4 \pi \sqrt{m_s m_e}} \rho,
\end{equation}
where $\rho$ is the liquid's density and $m_s$ the mass of the solvent molecule. For water at room temperature, this estimate is only about a factor of $5$ lower than the experimental value\cite{mehta_what_2025}.  Estimating  the density as $\rho \sim m_s/a^3$ and using $R\sim a$ in the Stokes formula, we obtain 
$$\gamma \sim \frac{3 \hbar }{2 a^2 } \sqrt{\frac{m_s} {m_e}}.$$
for the friction coefficient and 
\begin{equation} \label{velocity memory}
    \tau_{vel} \sim \frac{2 m a^2 }{3 \hbar } \sqrt{\frac{m_e} {m_s}} \sim \frac{a^2 \sqrt{m m_e} }{\hbar},
\end{equation}
for the velocity memory time. In the last estimate we have assumed that the particle's mass $m$ is of the same order of magnitude as $m_s$. 

We now introduce another important timescale describing diffusive dynamics:  the time it takes a diffusing particle to travel a distance comparable with its own size $R$:
\begin{equation} \label{diff time}
\tau_{diff } = \frac{R^2}{D} = 6\pi \eta \frac{R^3}{k_B T}.
\end{equation}
For a molecule of size $R=a$, using the TB viscosity estimate, we find
\begin{equation}
    \tau_{diff} \equiv \tau_{kin} \sim \frac{\hbar}{k_B T} \sqrt{\frac{m_s}{m_e}}=\nu_{Ey}^{-1}\sqrt{\frac{m_s}{m_e}}\gg \nu_{Ey}^{-1}.
\end{equation}
The significance of this time was noted by Mehta and Kondev\cite{mehta_what_2025}, who called it the {\it kinetic time}.  It is a curious product of the Eyring time $\nu_{Ey}^{-1}$ and  a large factor that is comparable to the square root of the proton and electron masses $\sqrt{m_p/m_e}$.  Comparing the kinetic time with the velocity memory time, Eq.~\ref{velocity memory}, we find
$$
\frac{\tau_{kin}}{\tau_{vel}} \sim \frac{E_{b}}{k_B T} = \mathcal{A}\gg 1,
$$
where $E_b$ was assumed to be comparable to the dissociation energy of the hydrogen atom, and where numerical constants of order 1 were dropped.  Thus the ratio of the two timescales is governed by the Arrhenius ratio, which is much greater than $1$. 

The fact that $\tau_{kin}$ is much greater than $\tau_{vel}$ is important: as in many biophysical problems we are concerned with the behavior of molecules at length scales greater than the molecular size $a$, it enables us to neglect inertial effects and to view molecular dynamics as purely diffusive. 
%even a microscopic particle forgets its own velocity much (a factor of $\mathcal{A}$) sooner than it takes  to diffuse a distance equal to its own dimensions is in argument in favor of neglecting inertial effects in solution dynamics, but one should keep in mind that whether the overdamped limit can be assumed is determined by the interplay of {\it three} timescales, $\tau_{Diff}$, $\tau_{vel}$, and inverse vibrational frequency $\omega^{-1}$. 

This assumption is even better justified  for large molecules (macromolecules such as proteins) whose overall dimension $R$ is much greater than the atomic size $a$. For those, the diffusion timescale, Eq.~\ref{diff time}, is proportional to the cube of the characteristic size, 
\begin{equation} \label{diff time scaling}
\tau_{diff} =\tau_{kin} \left(\frac{R}{a} \right)^3.
\end{equation}
One could imagine living in a world with protein molecules flying across cells like billiard balls, but the values of the fundamental physical constants make such a world unattainable.

\begin{figure}
    \centering
    \includegraphics[width=0.75\linewidth]{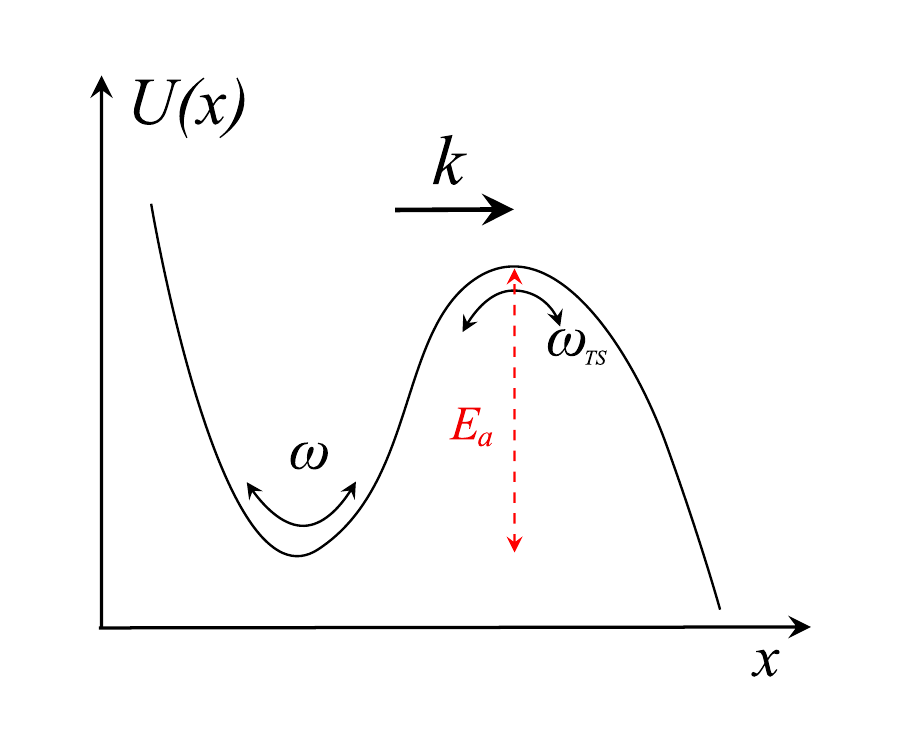}
    \caption{Kramers' model of a chemical reaction}
    \label{fig:Kramers}
\end{figure}

What is we apply the Langevin equation, Eq.~\ref{Langevin}, to a molecular bond?  This idea goes back to the classic paper by Kramers\cite{kramers_brownian_1940}, who proposed to use the Langevin equation as a description  of certain vibrational degrees of freedom participating in chemical reactions.  Kramers envisioned a chemical reaction as motion in a potential $U(x)$ describing a metastable state, with the particle stuck near its minimum for a while and finally escaping the minimum over the potential barrier whose height is $E_a$ (Eq.~\ref{Arrhenius}), as illustrated in Fig.~\ref{fig:Kramers}  

He found the rate of such a reaction to be given, again, by the Arrhenius law, Eq.~\ref{Arrhenius}, with the prefactor described by the formula
\begin{equation} \label{Kramers}
    \nu = \frac{\omega}{2\pi} \left[ \sqrt{1+\frac{\gamma^2}{4m^2\omega_{TS}^2}}-\frac{\gamma}{2 m \omega_{TS}}\right],
\end{equation}
where $\omega_{TS}$ is the upside-down barrier frequency, Fig.~\ref{fig:Kramers}. For our purposes, we can assume that $\omega\sim \omega_{TS}$. When $\gamma/m = \tau_{vel}^{-1}\ll \omega$, the dynamics in the potential $U(x)$ is underdamped, and Kramers' formula predicts $\nu \approx \omega/2\pi$, recovering the prefactor estimate of Eq.~\ref{prefactor}\footnote{Kramers also showed that Eq.~\ref{Kramers} becomes incorrect in the limit $\gamma\to 0$, but this regime is usually irrelevant for condensed-phase reactions.}  In the opposite limit of high friction, $\gamma/m=\tau_{vel}^{-1}\gg \omega$ one obtains
\begin{equation} \label{Kramers high friction}
\nu \approx \frac{m \omega  \omega_{TS}}{2\pi \gamma} \sim \frac{m\omega^2}{2\pi \gamma} =\frac{\kappa}{2\pi \gamma} =\frac{\omega^2 \tau_{vel}}{2\pi},    
\end{equation}
showing that the transition rate is inversely proportional to the friction coefficient. In this case, the rate coefficient $k$ is roughly the transition-state value multiplied by  a factor $\omega \tau_{vel}\ll 1$ . 

If we estimate the friction coefficient using the TB viscosity limit combined with the Stokes formula, Eq.~\ref{velocity memory}, we find that $\omega \tau_{vel}\sim 1$, since for the vibrational frequency, we have

\begin{equation} \label{damping estimate}
     \omega =\sqrt{\frac{\kappa}{m}}\sim ~\sqrt{\frac{E_b}{ma^2}}\sim \frac{\hbar}{a^2 \sqrt{m m_e} },
\end{equation}
where $E_b\sim \hbar^2/m_e a^2$  was used as a rough estimate of the bond energy. Therefore, according to our estimates, molecular vibrational degrees should be somewhere between underdamped and overdamped.  Experimentally, we know that molecular vibrations corresponding to bond stretches are much closer to underdamped. Indeed, the damping $\gamma/m$ is roughly comparable to the linewidth of the vibrational spectrum of a molecule; although spectral lines corresponding to bond stretches are significantly broader in the liquid phase than they are in the gas phase, the linewidth is typically only a fraction of the frequency $\omega$, making vibrational lines well-resolved (to a spectroscopist's delight). This discrepancy highlights inadequacy of using continuum hydrodynamics (i.e., the Stokes formula) as a description of microscopic energy dissipation in molecules. 

Trachenko-Brazhkin's microscopic estimate of viscosity is not the only one known. A different, much earlier estimate due to Eyring \cite{eyring_viscosity_1936} predicts, for the minimum possible viscosity (see Appendix):
$$
\eta=\eta_{Ey}\sim \frac{\eta_{TB}}{\mathcal{A}}
$$
Because Eyring's estimate is significantly lower  than the TB estimate, it will result in much weaker damping, $\gamma/m\ll \omega $ . Although this is more consistent with spectroscopy data, $\eta_{Ey}$ is at odds with the known bulk viscosity of water.  There seems to be no obvious reason to prefer $\eta_{Ey}$ to $\eta_{TB}$  when considering molecular vibrations, but the idea of using an {\it effective} (e.g., dependent on the frequency of motion -- see, e.g., ref.~\cite{elber_molecular_2020}) viscosity deserves consideration.    

If the assumption of overdamped dynamics fails for molecular vibrations, when is it justified? Since the estimate for $\tau_{vel}$ is essentially fixed by the quantum viscosity limit, molecular dimensions, and the mass, while the vibrational frequency $\omega$ depends on the energetic scale of molecular interactions as well as on the magnitude of displacements, softer molecular modes (smaller $\omega$), such as torsional vibrations, should be closer to the overdamped limit. Many biological phenomena are driven by ``entropic'' forces; in those cases the characteristic energies are comparable to $k_B T$, and the characteristic spring constants should be given not by Eq.~\ref{stiffness} but by
\begin{equation} \label{entropic stiffness} 
    \kappa \sim \frac{k_B T}{r^2},
\end{equation}
where $r$ is a characteristic length scale of the motion. Since $k_B T =E_b/\mathcal{A}\ll E_b$, the corresponding stiffness is nearly two orders of magnitude smaller than that of molecular vibrations, even if we assume $r\sim a$. In reality, we have $r\gg a$ for most biophysical phenomena of interest; as a result, the overdamped assumption is often justified when describing such phenomena. One of those, biomolecular folding, is discussed below. 

\subsection*{Speed limit of biomolecular folding}  There are many cellular phenomena involving a molecule undergoing diffusive search for another molecular target\cite{phillips_physical_2019}. The speed limit of those is determined by the target size and the diffusivity, which, in turn, is bounded by the quantum viscosity limit. Examples of such speed limits can be found in ref.~\cite{mehta_what_2025}. Here we will estimate the speed limit for protein folding, which is set by the molecular chain's diffusive search  in the conformational space. It is known that the rate of protein folding can be described by an Arrhenius law \cite{klimov_viscosity_1997,socci_diffusive_1996}, with the {\it free} energy barrier taking place of the activation energy in Eq.~\ref{Arrhenius}. The speed limit is the prefactor $\nu$. It could be argued that this cannot be greater than the inverse time it takes the unfolded polymer chain to significantly change its conformation\cite{thirumalai_time_1999}, which one may call the polymer's ``reconfiguration time''\cite{nettels_ultrafast_2007}. To estimate this time, we think of the unfolded protein as a ``random coil'' well studied by polymer theorists\cite{de_gennes_scaling_1979,grosberg_giant_1997}. The characteristic dimension $R$ of the chain can then be estimated by modeling the chain's backbone as a random walk in space, wherein the walker makes steps in random directions. Since the typical displacement of a random walker  grows as the square root of the number of steps, we expect 
$$
R\sim l  N^{1/2},
$$
where $l$ is a length scale associated with the step and $N$ is the number of monomers (polypeptide length). If we account for the fact that the backbone cannot cross itself, one gets a slightly different scaling law\cite{de_gennes_scaling_1979}, 
$$
R\sim lN^{3/5}.
$$
If we identify the length $l$  with the peptide bond length, which, in turn, is related to  the atomic scale $a$, we will then have succeeded in estimating $R$ from first principles, as long as we assume that $N$ is known\footnote{The minimum possible value of $N$ is arguably determined by energetic considerations, with small-$N$ chains being unstable thermodynamically - estimating $N$ from first principles is a difficult task that will not be attempted here}.  Setting $l=a$ will underestimate the actual value of $R$ because this does not account for the directional correlation between adjacent bonds and also underestimates the actual polypeptide bone length, but it still gives us the right order of magnitude. 

We could then define the chain reconfiguration time, $\tau_r=\nu^{-1}$,  as the time it takes two monomers (say chain ends)  to travel, via diffusion, across a distance comparable to $R$.  At first glance, predicting $\tau_r$ appears to be a difficult task, as the motion of the chain requires rearrangement of the backbone geometry  typically accomplished via hindered dihedral rotations. The timescale for such rotations will depend on the energetics, since a transition in the dihedral space involves an activation barrier. One can think of this rearrangement mechanism as overcoming ``internal friction'' or ``internal viscosity'' of the chain.  Fortunately, a theorem proven by Kuhn (see ref.~\cite{de_gennes_scaling_1979}) shows that, for long enough chains (i.e., large enough $N$), internal friction becomes negligible, and the overall chain reconfiguration time is controlled by the solvent viscosity. Moreover, the reconfiguration time can be estimated as the time it takes the entire chain, as a  whole,  to travel the distance $R$\footnote{A useful trick to explain this estimate is to imagine that the chain is divided in two halves, each with a diffusivity $\sim  D$,  having to travel a distance $\sim R$ relative to one another}, 
\begin{equation} \label{reconfig time}
\tau_r \sim \tau_{diff} \sim  \frac{R^2}{D} = 6\pi \eta \frac{R^3}{k_B T},
\end{equation}
where the Stokes formula was used to estimate the diffusivity of the chain. And while protein chains are not infinitely long, experiments\cite{soranno_quantifying_2012} and molecular simulations\cite{echeverria_concerted_2014,das_dynamics_2018} show that internal friction does not typically change the order of magnitude of the reconfiguration time. 

Another way to think of the speed limit of folding is using the Kramers high-friction limit, Eq.~\ref{Kramers high friction},  
$$
\nu \sim \frac{\kappa}{\gamma} 
$$
with $\kappa = k_B T/R^2$ for the entropic stiffness of the unfolded polymer (cf. Eq.~\ref{entropic stiffness}). Assuming that $\gamma$  is comparable to the Stokesian friction coefficient on an object of size $R$, one again finds $\nu \sim 1/\tau_r$, with $\tau_r$ given by Eq.~\ref{reconfig time}. 

Using the actual measured values of $R$ and the true water viscosity puts $\tau_r$ in a range of tens of nanoseconds for chains with $N<100$. Using instead our ``first-principles'' estimate for $R$ along with the TB quantum viscosity limit (cf. Eq.~\ref{diff time scaling}), we obtain 
\begin{equation} \label{reconfig time scaling}
\tau_{r} =\tau_{kin} \left(\frac{R}{a} \right)^3\sim \tau_{kin} N^{3/2}= \frac{\hbar}{k_B T} \sqrt{\frac{m_s}{m_e}} N^{3/2}
\end{equation}
(assuming $N^{1/2}$ scaling of the random walk length). For $N=100$ this gives a time of a few nanoseconds, about 1-2 orders of magnitude shorter than an expected experimental time\cite{kubelka_protein_2004}, because both the size $R$ and the actual water viscosity are underestimated and because factors like $2\pi$ were omitted. We have also ignored the entropic cost required for the chain to find the ``right'' conformations to initiate folding\cite{thirumalai_time_1999,makarov_topomer_2003}. Yet it is reassuring to know that the mere knowledge of the fundamental physical constants of Nature gets one this close to estimating the timescale of a rather complicated phenomenon.  

\begin{acknowledgments}
\noindent This work was supported by the US National Science Foundation. %Discussions with Graeme Henkelman are gratefully acknowledged.
\end{acknowledgments}

\appendix
\section{Eyring's microscopic theory for viscosity}

Eyring derives an estimate for molecular viscosity using (roughly) the following argument\cite{eyring_viscosity_1936}. Consider a layer of liquid of microscopic thickness $\sim a$ subjected to a shear stress $f$ (force per unit area). We have  
\begin{equation} \label{viscosity law}
f=\eta v/a
\end{equation}
where  $v/a$ is the shear rate. The presence of the force makes molecules more likely to hop in one direction than in the other. Calling $k_{+}(f)$ and $k_{-}(f)$ the hopping rates in the two directions, thermodynamic considerations result in $k_{+}(f)/k_{-}(f)= e^{fa^3/k_B T}$ , where $f a^2 \times a$ is the free energy change upon a microscopic ``hop'' of length $a$.  Setting then  $k_{\pm}(f)=k_0  e^{\pm fa^3/2 k_B T}$, where $k_0$ is the hopping rate in the absence of the force, we estimate the microscopic drift velocity of molecules hopping preferentially along the force as
$$
v=a [k_{+}(f)-k_{-}(f)] \approx \frac{k_0 a^4 f}{k_B T}
$$
where we assumed that $fa^3\ll k_B T$. Comparing this with Eq.~\ref{viscosity law}, we obtain
$$
\eta =\frac{k_B T}{k_0 a^3},
$$
The hopping rate $k_0$ obeys the Arrhenius law, Eq.~\ref{Arrhenius}, giving the viscosity the expected exponential dependence on the activation barrier. The lowest possible viscosity value is obtained if one sets
$$
k_0=\nu=\frac{\omega}{2 \pi},
$$
which results in 
$$
\eta =\eta_{Ey}=\frac{2 \pi k_B T}{a^3}.
$$
Comparing this with the Trachenko-Brazhkin viscosity, it is straightforward to verify that 
$$
\frac{\eta_{Ey}}{\eta_{TB}} \sim \frac{1}{\mathcal{A}},
$$
where numerical factors have been omitted. 
\bibliographystyle{ieeetr}
\bibliography{references.bib}
\end{document}